\documentclass[aps,prl,showpacs,twocolumn,superscriptaddress]{revtex4} %%% revtex4 (print form)
\newcommand{\bfi}[1]{\mbox{\boldmath $#1$}}
\newcommand{\Lower}[1]{\smash{\lower 1.5ex \hbox{#1}}}
\newcommand{\LN}{$\Lambda N$}

\newcommand{\LL}{$\Lambda\Lambda$}
\newcommand{\SN}{$\Sigma N$}
\newcommand{\LS}{$\Lambda\Sigma$}
\newcommand{\NX}{$N\Xi$}
\newcommand{\SigSig}{$\Sigma\Sigma$}
\newcommand{\LNSN}{$\Lambda N-\Sigma N$}
\newcommand{\LLNX}{$\Lambda\Lambda-N\Xi$}

\newcommand{\LLNXSS}{$\Lambda\Lambda-N\Xi-\Sigma\Sigma$}
\newcommand{\BL}{$B_\Lambda$}
\newcommand{\BLL}{$B_{\Lambda\Lambda}$}
\newcommand{\DBLL}{$\Delta B_{\Lambda\Lambda}$}

\newcommand{\HII}{$^2$H}
\newcommand{\HIII}{$^3$H}
\newcommand{\HeIII}{$^3$He}
\newcommand{\HeIV}{$^4$He}

\newcommand{\HIIIL}{$_\Lambda^3$H}

\newcommand{\HIVL}{$_\Lambda^4$H}

\newcommand{\HeIVL}{$_\Lambda^4$He}

\newcommand{\HIVLs}{$_\Lambda^4$H$^\ast$}
\newcommand{\HeIVLs}{$_\Lambda^4$He$^\ast$}
\newcommand{\HeVL}{$_\Lambda^5$He}
\newcommand{\HIVLL}{$_{\Lambda\Lambda}^{\ \ 4}$H}

\newcommand{\HVLL}{$_{\Lambda\Lambda}^{\ \ 5}$H}
\newcommand{\HVX}{$_{\Xi}^{5}$H}
\newcommand{\HeVX}{$_{\Xi}^{5}$He}
\newcommand{\HVLS}{$_{\Lambda\Sigma}^{\ \ 5}$H}
\newcommand{\HVSS}{$_{\Sigma\Sigma}^{\ \ 5}$H}
\newcommand{\HeVLL}{$_{\Lambda\Lambda}^{\ \ 5}$He}
\newcommand{\HeVILL}{$_{\Lambda\Lambda}^{\ \ 6}$He}

\usepackage{graphicx}%%% Does not work in `documentstyle' env

\begin{document} 				%%% revtex4

\title{ 
Full-Coupled Channel Approach to Doubly Strange $s$-Shell Hypernuclei 
}
\author{H.~{Nemura}} 		%%% Here is for revtex4
\affiliation{ 					%%% Here is for revtex4
Institute of Particle and Nuclear Studies, KEK, Tsukuba 305-0801, Japan
} 						%%% Here is for revtex4
\author{S.~{Shinmura}}
\affiliation{ 					%%% Here is for revtex4
Department of Information Science, Gifu University,  Gifu 501-1193, Japan
} 						%%% Here is for revtex4
\author{Y.~{Akaishi}} 		%%% Here is for revtex4
\affiliation{ 					%%% Here is for revtex4
College of Science and Technology, Nihon University, Funabashi 274-8501, Japan
} 						%%% Here is for revtex4
\author{Khin~Swe~Myint} 		%%% Here is for revtex4
\affiliation{ 					%%% Here is for revtex4
Department of Physics, Mandalay University, Mandalay, Union of Myanmar
} 						%%% Here is for revtex4

\date{\today}

\begin{abstract} 				%%% revtex
 We describe {\it ab initio} calculations of doubly strange, $S=-2$, 
 $s$-shell hypernuclei (\HIVLL, \HVLL, \HeVLL\ and \HeVILL) 
 as a first attempt to explore the few-body problem of the 
 {\it full}-coupled channel scheme 
 for these systems. 
 The wave function includes \LL, \LS, \NX\ and \SigSig\ channels. 
 Minnesota $NN$, D2$^\prime$ $YN$, 
 and simulated $YY$ potentials based on the Nijmegen hard-core model, 
 are used. 
 Bound-state solutions of these systems 
 are obtained. 
 We find that a set of phenomenological $B_8B_8$ 
 interactions among the octet baryons in $S=0, -1$ and $-2$ sectors, 
 which is consistent with all of the available experimental binding 
 energies of $S=0, -1$ and $-2$ $s$-shell (hyper-)nuclei, 
 can predict a particle stable bound-state of \HIVLL. 
 For \HVLL\ and \HeVLL, 
 \LNSN\ and $\Xi N-\Lambda\Sigma$ 
 potentials 
 enhance the net \LLNX\ coupling, 
 and a large $\Xi$ probability 
 is obtained even for a weaker \LLNX\ 
 potential. 
\end{abstract} 				%%% revtex

 %%% revtex4

\maketitle 					%%% ptptex \& revtex4

Both 
recent 
experimental and theoretical studies of doubly strange ($S=-2$) 
$s$-shell hypernuclei (\HIVLL, \HVLL, \HeVLL, and \HeVILL) 
are 
the 
utmost 
exciting topics in the field of 
hypernuclei\cite{Nagara,Ahn,Kumagai,FG,Nem03,Kahana,KSMSA,FGS,LanskYama,Afnan,Yamada}. 
An 
experimental report\cite{Nagara} on a new observation of \HeVILL\ 
has had 
a significant 
impact 
on strangeness nuclear physics. 
The {\it Nagara} event 
provides unambiguous identification of \HeVILL\ production, and suggests that 
the \LL\ interaction strength is rather weaker than that expected from an 
older experiment\cite{Prowse}. 

The BNL-AGS E906 experiment\cite{Ahn} has conjectured 
a formation of \HIVLL, 
in accordance with our earlier predictions\cite{Nakaichi,Nem00} that 
\HIVLL\ would exist as 
a particle stable bound-state against strong decay. 
If this is the case, the \HIVLL\ would be the lightest bound state among 
doubly strange hypernuclei. 
However, a theoretical study\cite{Kumagai} of the weak-decay modes 
from \HIVLL\ does  
not support this conjecture, 
and our earlier 
studies should be reanalyzed by taking account of 
the new datum, Nagara event. 

A recent Faddeev-Yakubovsky search for \HIVLL\cite{FG} found no 
bound-state solution over a wide range of \LL\ interaction strengths, 
although this conclusion has been in conflict with the result 
calculated by authors using a variational method\cite{Nem03}. 
The total binding energy is more sensitive to the $^3S_1$ channel of the 
\LN\ interaction than to the $^1S_0$ \LL\ interaction, because the 
number of the $^3S_1$ \LN\ pairs is three times larger than the 
number of the $^1S_0$ \LL\ pair, 
as was discussed 
in Ref.~\cite{FG}. 
Therefore, the spin-dependent part of the \LN\ interaction has to be 
determined very carefully. 
The algebraic structure of the 
$({\bfi \sigma}_\Lambda\cdot{\bfi \sigma}_N)$ interaction for the $S=-2$ 
systems is similar to the structure for the \HeVL. 
Namely, the \LN\ interaction, which is utilized in the theoretical 
search for \HIVLL, has to reproduce the experimental \BL(\HeVL) as well 
as the \BL's of $A=3, 4$ $S=-1$ hypernuclei. 
However, there is a long-standing problem known as the 
\HeVL\ anomaly\cite{survey}, since the publication by Dalitz {\it et al.} 
in 1972\cite{DHT}. 
Recently, Akaishi {\it et al.}\cite{Akaishi} successfully resolved 
the anomaly 
by explicitly taking account of \LNSN\ coupling.

Considering the fact that the \LL\ system couples to the \NX\ and 
$\Sigma\Sigma$ states, and also the \LN\ system couples to the 
\SN\ states, 
a theoretical search for \HIVLL\ should be made in a 
fully coupled channel formulation with a set of interactions 
among the octet baryons. 
The \HVLL-\HVX\ (or \HeVLL-\HeVX) mixing due to 
$\Lambda\Lambda\!\!-\!\!N\Xi$ 
coupling is also interesting topic, 
since the $\alpha$-formation effect could be 
significant\cite{KSMSA,LanskYama}. 
Thus, the purpose of this study is threefold: 
First is to describe a systematic study 
for the complete set of $s$-shell hypernuclei with 
$S\!\!=\!\!-\!2$~in~a~framework of a full-coupled channel formulation. 
Second is to make a conclusion if a set of baryon-baryon interactions, 
which is consistent with the experimental data, 
predicts a particle stable bound state of \HIVLL. 
The third is to explore the fully hyperonic mixing of \HVLL, 
including the 
$\Lambda N\!\!-\!\!\Sigma N$ 
transition potential 
in addition~to~$\Lambda\Lambda\!\!-\!\!N\Xi\!\!-\!\!\Sigma\Sigma$.

The wave function of a system with $S=-2$, 
comprising $A(=N+Y)$ octet baryons, 
has four isospin-basis components. 
For example, \HeVILL\ has four components as $ppnn\Lambda\Lambda$, 
$NNNNN\Xi$, $NNNN\Lambda\Sigma$ and $NNNN\Sigma\Sigma$. 
We abbreviate these components as \LL, \NX, \LS\ and $\Sigma\Sigma$, 
referring to the last two baryons. 
The Hamiltonian of the system is hence given by $4\times 4$ components 
as 
\begin{equation}
 H=\left(\begin{array}{cccc}
    H_{\Lambda\Lambda} & V_{N\Xi-\Lambda\Lambda} & 
     V_{\Lambda\Sigma-\Lambda\Lambda} & V_{\Sigma\Sigma-\Lambda\Lambda} \\
	  V_{\Lambda\Lambda-N\Xi} & H_{N\Xi} & 
	   V_{\Lambda\Sigma-N\Xi} & V_{\Sigma\Sigma-N\Xi} \\
	  V_{\Lambda\Lambda-\Lambda\Sigma} & V_{N\Xi-\Lambda\Sigma} & 
	   H_{\Lambda\Sigma} & V_{\Sigma\Sigma-\Lambda\Sigma} \\
	  V_{\Lambda\Lambda-\Sigma\Sigma} & V_{N\Xi-\Sigma\Sigma} &
	   V_{\Lambda\Sigma-\Sigma\Sigma} & H_{\Sigma\Sigma} \\
\end{array}
\right),
\label{HTOT}
\end{equation}
where $H_{B_1B_2}$ operates on the $B_1B_2$ component, 
and $V_{B_1B_2-B_1^\prime B_2^\prime}$ is the sum of 
all possible two-body transition potential connecting the 
$B_1B_2$ and $B_1^\prime B_2^\prime$ components: 
\begin{eqnarray}
 V_{\Lambda\Lambda-N\Xi} &=& v_{\Lambda\Lambda-N\Xi}, \label{LLNX}\\
 V_{\Lambda\Lambda-\Lambda\Sigma}
  &=& \sum_{i=1}^N v_{N_i\Lambda-N_i\Sigma}, \quad %\\
 V_{N\Xi-\Lambda\Sigma} = v_{N\Xi-\Lambda\Sigma}, \label{LLLS}\\
 V_{\Lambda\Lambda-\Sigma\Sigma} &=& v_{\Lambda\Lambda-\Sigma\Sigma}, \quad %\\
 V_{N\Xi-\Sigma\Sigma} = v_{N\Xi-\Sigma\Sigma}, \label{LLSS}\\
 V_{\Lambda\Sigma-\Sigma\Sigma} &=& \sum_{i=1}^N v_{N_i\Lambda-N_i\Sigma}
  + v_{\Lambda\Sigma-\Sigma\Sigma}. 
\end{eqnarray}
Note that we take account of {\it full}-coupled channel potentials 
including the $\Lambda N\!\!\!-\!\!\Sigma N$ 
and $N\Xi\!\!-\!\!\Lambda\Sigma$ transitions 
(Eq.(\ref{LLLS})) 
in the $^3\!S_1$ channel, 
while other full-coupled channel 
approaches 
(e.g., Refs.\cite{Afnan,Yamada}) 
only take $\Lambda\Lambda\!\!-\!\!N\Xi\!\!-\!\!\Sigma\Sigma$ 
in the $^1\!S_0$ channel (Eqs.(\ref{LLNX}) and (\ref{LLSS})) 
into account.

In the present calculations, we use the Minnesota potential\cite{Minn} 
for the $NN$ interaction and D2$^\prime$ for the $YN$ interaction. 
The Minnesota potential reproduces reasonably well both the binding 
energies and sizes of few-nucleon systems, 
such as \HII, \HIII, \HeIII\ and \HeIV\cite{PreciseVS}. 
The D2$^\prime$ potential is a modified potential from the original D2 
potential\cite{Akaishi}. 
The strength of the long-range part 
($V_b$ in Table I of Ref.~\cite{Akaishi}) of 
the D2$^\prime$ potential in the \LN-\LN\ $^3S_1$ 
channel is reduced by multiplying by a factor ($0.954$) 
in order to reproduce the experimental \BL(\HeVL) value. 
The calculated \BL\ values for the $\Lambda$ hypernuclei 
(\HIIIL, \HIVL, \HeIVL, \HIVLs, \HeIVLs, and \HeVL) are 
$0.056$, $2.23$, $2.17$, $0.91$, $0.89$, and $3.18$ MeV, respectively. 
For the $YY$ interaction, we use a full-coupled channel potential among 
the octet baryons in both the spin triplet and the spin singlet channels. 
We assume that the $YY$ potential consists of only the central 
component, 
and the effect due to the non-central force (e.g., tensor force) 
should be included into the central part effectively. 
The $YY$ potential has Gaussian form factors, whose parameters 
are set to 
reproduce the low-energy $S$ matrix of the Nijmegen hard-core model 
D (ND) 
or F (NF)\cite{ND77NF79}. 
We take the hard-core radius to be 
$r_c=0.56271(0.44915)$ 
fm in the 
spin singlet (triplet) channel for the ND, 
whereas 
$r_c=0.52972(0.52433)$ 
fm is used in the singlet (triplet) channel 
for the NF. 
Each number is the same as the hard-core radius of the $YN$ sector in 
each channel for each model. 
The strength parameters are firstly determined on a charge basis, 
and then the strength parameters on an isospin basis are constructed 
from the charge-basis parameters. 
We denote ND$_S$ (NF$_S$) for the simulating ND (NF) potential.

The calculations are made by using the 
stochastic variational method\cite{Kukulin,SVM}. 
This is essentially 
along the lines of 
Ref.~\cite{abinitio}, 
except for the isospin function. 
The isospin function consists of four components, in accordance with 
Eq.~(\ref{HTOT}). 
The reader is referred to Refs.~\cite{SVM,abinitio} for the details of 
the method.

\begin{figure*}[]
 \includegraphics[width=1.\textwidth]{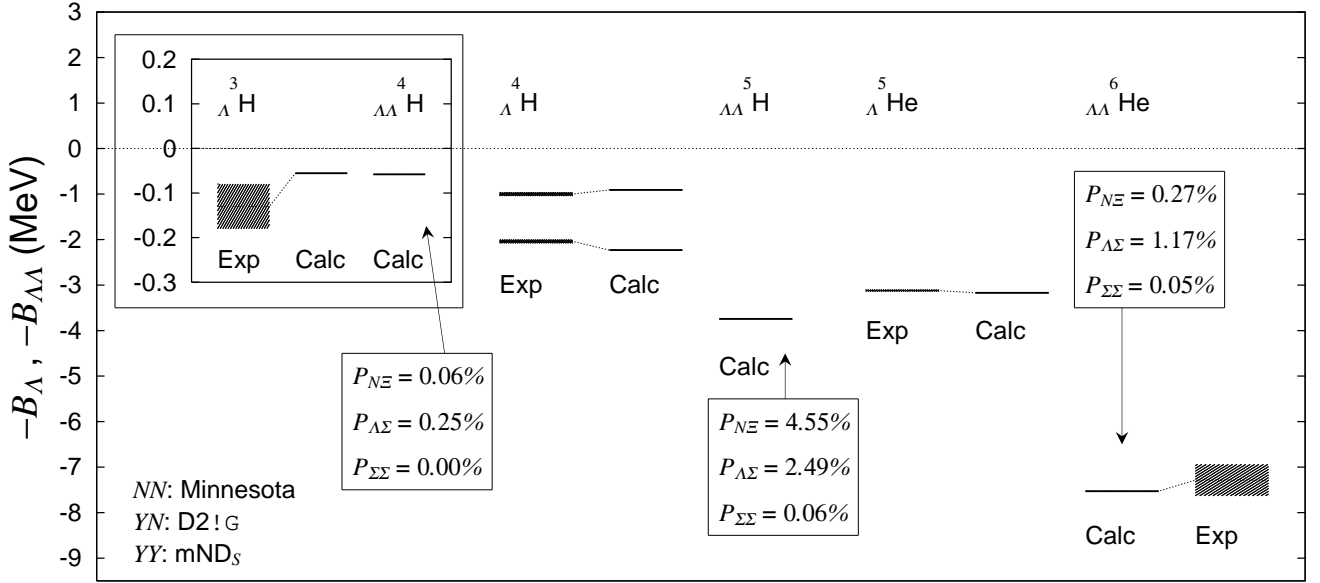}
 \caption{$\Lambda$ and \LL\ separation energies of $A=3-6$, $S=-1$ and 
 $-2$ $s$-shell hypernuclei. 
 The Minnesota $NN$, D2$^\prime$ $YN$ and mND$_S$ $YY$ potentials are 
 used. 
 The width of the line for the experimental \BL\ or \BLL\ value 
 indicates the experimental errorbar. 
 The probabilities of the \NX, \LS\ and $\Sigma\Sigma$ components are 
 also shown for the \LL\ hypernuclei. 
 \label{BEA3456wNDredLLwF0.8} }
\end{figure*}
\begin{table}[]
 \caption{\LL\ separation energies, given in units of MeV, 
 of $A=4-6$, $S=-2$ $s$-shell hypernuclei. 
 \label{BEA456}}
 \begin{ruledtabular}
  \begin{tabular}{lcccc}
   $YY$ & \BLL(\HIVLL) & \BLL(\HVLL) & \BLL(\HeVLL) & \BLL(\HeVILL) \\
   \hline
   ND$_S$  & $0.107$ & $4.04$ & $3.96$ & $7.93$ \\
   mND$_S$ & $0.058$ & $3.74$ & $3.66$ & $7.53$ \\
   NF$_S$  & $0.127$ & $3.84$ & $3.77$ & $7.52$ \\
   Exp & & & & $7.25\pm0.19^{+0.18}_{-0.11}$ \\ 
  \end{tabular}
 \end{ruledtabular}
\end{table}

Table~\ref{BEA456} lists the \BLL\ values for $S=-2$ hypernuclei. 
Using ND$_S$ or NF$_S$ $YY$ potential, we have obtained the bound-state 
solutions of \HIVLL, \HVLL, \HeVLL\ and \HeVILL. 
In the case of ND$_S$ $YY$ potential, 
we have 
\begin{eqnarray}
\Delta B_{\Lambda\Lambda}^{\mbox{(calc)}}(\mbox{\HeVILL})
&=&B_{\Lambda\Lambda}^{\mbox{(calc)}}(\mbox{\HeVILL})
-2B_{\Lambda}^{\mbox{(calc)}}(\mbox{\HeVL})\nonumber \\
&=&1.58 \mbox{MeV}, 
\end{eqnarray}
which is slightly larger than the experimental value 
($\Delta B_{\Lambda\Lambda}^{\mbox{(exp)}}(\mbox{\HeVILL})=1.01\pm
0.20^{+0.18}_{-0.11}$ MeV\cite{Nagara}), 
while the NF$_S$ reproduces the experimental value fairly well. 
We have also calculated these systems using a modified ND$_S$ (mND$_S$) 
$YY$ potential; 
The strength of the \LL\ diagonal part of the mND$_S$ potential is 
reduced by multiplying by a factor of $0.8$ from the original ND$_S$ 
in order to reproduce the experimental \DBLL(\HeVILL). 
The scattering length and effective range parameters for the ND$_S$, 
mND$_S$, and NF$_S$ are (given in units of fm): 
$(a_s,r_s)=(-1.37, 4.98)$, $(-0.91, 6.25)$, and $(-0.40, 12.13)$, 
respectively. 
The scattering length for the mND$_S$ or NF$_S$ is consistent with the 
other analyses\cite{FG,Afnan} concerning the Nagara event. 
We should note that the 
mND$_S$ 
potential 
predicts 
the particle stable bound state of \HIVLL; The obtained energy is very 
close to, but still ($0.02$~MeV) lower than, the $\mbox{\HIIIL}+\Lambda$ 
threshold. 
Therefore, due to the result for mND$_S$ or NF$_S$, 
we should come to the following novel conclusion: 
A set of phenomenological baryon-baryon 
interactions among the octet baryons in $S=0, -1$ and $-2$ sectors, 
which is consistent with 
the Nagara event as well as 
all the experimental binding energies of 
$S=0$ and $-1$ $s$-shell (hyper-) nuclei, 
can predict a particle stable bound state of \HIVLL. 

Figure~\ref{BEA3456wNDredLLwF0.8} schematically displays the 
present results of the full-coupled channel calculations 
of $A=3-6, S=-1, -2$ hypernuclei,  
using the mND$_S$. 
Since the present calculation has been made on the isospin basis, 
the results for \HeVLL\ are qualitatively similar to the results for 
\HVLL, so that we omit the explicit result for \HeVLL. 
Fig.~\ref{BEA3456wNDredLLwF0.8} 
also displays the probabilities of the \NX, \LS\ 
and \SigSig\ components for the $S=-2$ hypernuclei. 
In the case of NF$_S$, the probabilities are (given in percentage): 
$(P_{N\Xi},P_{\Lambda\Sigma},P_{\Sigma\Sigma})=(0.58, 0.37, 0.03)$ 
for \HIVLL, 
$(3.10,2.10,0.10)$ for \HVLL, and 
$(1.33,1.13,0.10)$ for \HeVILL, respectively. 
In the present calculations, the \LL\ component is the main part of the 
wave function. 
No unrealistic bound states were found for the $YY$ subsystem, 
since the hard-core model hardly 
incorporates an unrealistic strong attractive force in the short-range 
region, in contrast to the soft core model, such as NSC97e or 
NSC97f\cite{Yamada}. 
This is one of the reasons why we 
used the $YY$ potential constructed from the hard-core model, 
for 
the first attempt to the full-coupled channel calculation. 
$P_{\Sigma\Sigma}$'s are very small 
for both systems, 
due to a large mass difference between the \LL\ and \SigSig\ channels 
($m_{\Sigma\Sigma}-m_{\Lambda\Lambda}\cong 155$~MeV). 

We should emphasize that 
the $P_{N\Xi}$(\HVLL) obtained by the mND$_S$ 
has a surprisingly large value ($4.55$\%), 
which is larger than the $P_{N\Xi}$(\HVLL) 
obtained by the NF$_S$ ($3.10$\%), 
in spite of the fact that 
the strength of the \LLNX\ coupling 
potential of the ND is rather weaker than 
that of the NF. 
This does not imply that a stronger \LLNX\ coupling 
potential 
means a larger $P_{N\Xi}$ probability. 
This is in 
remarkable contrast with other calculations based on 
$(t\!+\!\Lambda\!+\!\Lambda)$ and $(\alpha\!+\!\Xi^-)$ 
two-channel model\cite{KSMSA,LanskYama}.

Although the present calculation assumes no simplified structures, 
such as $(t\!+\!\Lambda\!+\!\Lambda)$ and $(\alpha\!+\!\Xi^-)$, 
this kind of model is useful to make a clear explanation of the 
complicated full coupling dynamics of the $A\!\!=\!\!5, S\!\!=\!\!-2$ hypernucleus. 
Let us consider a set of simple $core\ nucleus+\!Y(+Y)$ model wave 
functions for the~\HVLL: 
\begin{eqnarray}
 &&|\mbox{\HVLL}\rangle =
  \psi_t\times\psi_{\Lambda\Lambda}\times\psi_{\Lambda\Lambda-t},\\
 &&|\mbox{\HVX}\rangle =
  \psi_\alpha\times\psi_{\Xi^-}\times\psi_{\Xi^--\alpha},\label{WFalphX}\\
 &&|\mbox{\HVLS}\rangle_{S_{\Lambda\Sigma}}=
  \sqrt{1\over 3}\left[\psi_t\times[\psi_{\Lambda\Sigma^0}]_{S_{\Lambda\Sigma}}\right]
  \times\psi_{\Lambda\Sigma^0-t}
  \nonumber\\
 &&\qquad\qquad\qquad
  -\sqrt{2\over 3}\left[\psi_h\times[\psi_{\Lambda\Sigma^-}]_{S_{\Lambda\Sigma}}\right]
  \times\psi_{\Lambda\Sigma^--h}\nonumber\\
 &&\qquad\qquad\qquad\qquad\qquad\qquad
  (\mbox{for}\ S_{\Lambda\Sigma}=0\ \mbox{or}\ 1),%\\
\end{eqnarray}
where $\psi_c\ (c=t,h,\alpha)$ is the wave function (WF) of the core 
nucleus, 
$\psi_{YY}\ (YY=\Lambda\Lambda,\Xi^-,\Lambda\Sigma)$ is the WF 
of the hyperon(s), and 
$\psi_{YY-c}$ is the WF that describes the relative motion 
between $YY$ and $c$. 
We assume that all of the baryons occupy the same $(0s)$ orbit. 
For the \HVLS\ state, 
we have two independent states for the WF 
$\psi_{\Lambda\Sigma}$, 
that the spin of two hyperons ($S_{\Lambda\Sigma}$) is 
either a singlet or a triplet. 
Since the \SigSig\ component plays a minor role, 
we omit the \HVSS\ state. 
Using these WFs, we can obtain the algebraic factors for each averaged 
coupling potential of the allowed spin state, $\bar{v}^s$ or $\bar{v}^t$:
\begin{eqnarray}
 \langle V_{\Lambda\Lambda-N\Xi}\rangle
  \!\!\! &=& \!\!\! \sqrt{1\over 2} \bar{v}^s_{\Lambda\Lambda-N\Xi},\\ % \nonumber \\
 \langle V_{\Lambda\Lambda-\Lambda\Sigma} \rangle
  \!\!\! &=& \!\!\! \left\{\!\!\!
       \begin{array}{ll}
	\sqrt{9\over 8}\bar{v}^t_{N\Lambda-N\Sigma}
	 +\sqrt{1\over 8}\bar{v}^s_{N\Lambda-N\Sigma}
	 & (\mbox{for\ }S_{\Lambda\Sigma}\!=\!0 ),\\
	\sqrt{3\over 8}\bar{v}^t_{N\Lambda-N\Sigma}
	 -\sqrt{3\over 8}\bar{v}^s_{N\Lambda-N\Sigma}
	 & (\mbox{for\ }S_{\Lambda\Sigma}\!=\!1 ),
       \end{array}\right.\\ % \nonumber \\
 \langle V_{N\Xi-\Lambda\Sigma}\rangle
  \!\!\! &=& \!\!\! \left\{
       \begin{array}{ll}
	-\sqrt{3\over 4}\bar{v}^s_{N\Xi-\Lambda\Sigma}
	 & (\mbox{for\ }S_{\Lambda\Sigma}=0),\\
	{3\over 2}\bar{v}^t_{N\Xi-\Lambda\Sigma}
	 & (\mbox{for\ }S_{\Lambda\Sigma}=1).
       \end{array}\right. %\nonumber %\\
\end{eqnarray}
The 
$v_{\Lambda\Lambda-N\Xi}$ potential 
is suppressed by 
a factor of $\sqrt{1/2}$ for the $A=5$ hypernucleus. 
The 
$v_{N\Lambda-N\Sigma}$ 
and $v_{N\Xi-\Lambda\Sigma}$ 
potentials, 
particularly in the spin triplet channel, 
play significant roles instead. 
Namely, 
these equations imply that 
the $\Lambda\Sigma$ component strongly couples 
both to the \LL\ and to the \NX\ components, and 
the $\Lambda\Sigma$ component plays a crucial role in the hypernucleus.

The normalized energy expectation 
values of the Hamiltonian~(\ref{HTOT}) for \HVLL\ are 
(given in units of MeV), 

\noindent
\begin{eqnarray}
 h&=&\left(
  \begin{array}{ccc}
   \langle H_{\Lambda\Lambda}\rangle\over P_{\Lambda\Lambda} &
    \langle V_{N\Xi-\Lambda\Lambda}\rangle\over\sqrt{P_{\Lambda\Lambda}P_{N\Xi}} &
    \langle V_{\Lambda\Sigma-\Lambda\Lambda}\rangle\over\sqrt{P_{\Lambda\Lambda}P_{\Lambda\Sigma}}\\
   \langle V_{\Lambda\Lambda-N\Xi}\rangle\over\sqrt{P_{\Lambda\Lambda}P_{N\Xi}} &
    \langle H_{N\Xi}\rangle\over P_{N\Xi} &
    \langle V_{\Lambda\Sigma-N\Xi}\rangle\over\sqrt{P_{N\Xi}P_{\Lambda\Sigma}}\\
   \langle V_{\Lambda\Lambda-\Lambda\Sigma}\rangle\over\sqrt{P_{\Lambda\Lambda}P_{\Lambda\Sigma}} &
    \langle V_{N\Xi-\Lambda\Sigma}\rangle\over\sqrt{P_{N\Xi}P_{\Lambda\Sigma}} &
    \langle H_{\Lambda\Sigma}\rangle\over P_{\Lambda\Sigma}
  \end{array}
\right)\nonumber\\
 &=&\left\{
     \begin{array}{l}
      \left(
       \begin{array}{ccc}
	-9.12 & -1.82 & -14.52 \\
	-1.82 &  5.02 & -10.38 \\
	-14.52 & -10.38 & 92.44
       \end{array}
     \right)\quad (\mbox{for the mND}_S),\\
      ~\\
      \left(
       \begin{array}{ccc}
	-6.10 & -20.49 & -14.92 \\
	-20.49 & 115.4 & -10.01 \\
	-14.92 & -10.01 & 101.62 
       \end{array}
     \right)\quad (\mbox{for the NF}_S).
     \end{array}
\right.%\\
\end{eqnarray}
Here, we display only the $3\times 3$ components of the Hamiltonian 
(\ref{HTOT}), comprising \LL, \NX, and \LS, 
since the contributions from the \SigSig\ component are not large. 
If we solve the eigenvalue problem, $\det(h-\lambda I)\!=\!0$, 
we obtain the ground state energy, $E\!=\!-11.82$ MeV ($-11.82$ MeV), 
and the probability, $P_{N\Xi}\!=\!3.99\%$ (2.83\%), for the mND$_S$ 
(NF$_S$).  
The first $2\!\times\! 2$ components for the mND$_S$ are quite different 
from those 
for the NF$_S$, 
while the last row and the last column are qualitatively similar to each 
other. 
If we solve the eigenvalue problem of only the first 
$2\!\times\! 2$ subspace, including the \LL\ and the \NX, 
we obtain the ground state energy, $E\!=\!-9.35$ MeV ($-9.46$ MeV), and the 
probability, $P_{N\Xi}\!=\!1.57\%$ ($2.62\%$), for the mND$_S$ (NF$_S$). 
This clearly means that the couplings between the (\LL, \NX) and \LS\ 
components have to be taken into account %. 
in order to obtain an accurate solution. 
In the case of mND$_S$, the large coupling potentials, 
$\langle V_{\Lambda\Lambda-\Lambda\Sigma}\rangle$ and 
$\langle V_{N\Xi-\Lambda\Sigma}\rangle$, also 
enhance the 
$P_{N\Xi}$ probability. 
On the other hand, 
for NF$_S$, 
these coupling potentials hardly enhance $P_{N\Xi}$, though the total 
binding energy significantly increases. 
The large repulsive energy in the \NX\ component, which mainly comes 
from the 
$N\Xi\!-\!N\Xi$ 
diagonal potential in the $I\!\!=\!\!0$, $^1\!S_0$ channel, 
suppresses the net effect of the $\Lambda\Lambda\!-\!N\Xi$ 
coupling for~NF$_S$. 
On the other hand, the $N\Xi\!-\!N\Xi$ potential of mND$_S$ is weakly 
attractive, 
which is consistent with recent experimental 
data\cite{E885,Nakazawa,E224}.

In summary, we have performed full-coupled channel {\it ab initio} 
calculations for the complete set of doubly strange $s$-shell 
hypernuclei. 
Two kinds of $YY$ interactions, mND$_S$ and NF$_S$, reproduce the 
\DBLL(\HeVILL) of the Nagara event. 
We obtained bound-state solutions for \HIVLL, \HVLL\ and \HeVLL\ by 
using these $YY$ interactions. 
We thus conclude that a set of phenomenological $B_8B_8$ 
interactions among the octet baryons in $S=0, -1, -2$ sectors, 
which is consistent with all of the available experimental binding 
energies of the $S=0,-1,-2$ $s$-shell (hyper-) nuclei, can predict 
a particle stable bound state of \HIVLL. 
For the \HVLL\ (and \HeVLL), 
the probability $P_{N\Xi}$ by using the mND$_S$ is larger than 
the $P_{N\Xi}$ by using the NF$_S$. 
We found that the \LNSN\ and $N\Xi-\Lambda\Sigma$ potentials 
make a larger $P_{N\Xi}$(\HVLL) for the weak \LLNX\ %potential 
and attractive $N\Xi-N\Xi$ potentials 
of mND$_S$, 
whereas the net effect of the stronger \LLNX\ coupling potential of 
NF$_S$ is suppressed 
in \HVLL\ due to the 
repulsive $N\Xi-N\Xi$ 
potential in the $I=0$, $^1$S$_0$ 
channel. 
The one-boson-exchange potential models for the $B_8B_8$ interactions 
have sound bases of the $SU(3)$ symmetry 
and are widely accepted, 
though uncertainties of the interactions in the $S=-2$ sector are still 
large because of limitation of experimental information. 
The NSC97 models have a crucial defect in \LLNXSS\ couplings. 
Therefore, 
we have attempted only two possible cases with the models ND and NF, 
which have different characters in the $S=-2$ sector 
(strengths of \LLNX\ coupling and $N\Xi-N\Xi$ diagonal potentials). 
We do hope that a future experimental facility (e.g., J-PARC) develops 
our knowledge of the $S=-2$ interactions.

\begin{acknowledgments}         %%% revtex
The authors are thankful to K.~Nakazawa for useful communications. 
The calculations were made using the RCNP's SX-5 computer 
and the KEK SR8000 computer. 
\end{acknowledgments}          %%% revtex

\end{document}